# Role of salt valency in the switch of H-NS proteins between DNA-bridging and DNA-stiffening modes

Marc Joyeux[(#)]

*Laboratoire Interdisciplinaire de Physique,*

*CNRS and Université Grenoble Alpes,*

*Grenoble, France*

**Running title**: bridging/stiffening modes of H-NS

**Abstract:** This work investigates the interactions of H-NS proteins and bacterial genomic DNA through computer simulations performed with a coarse-grained model. The model was developed specifically to study the switch of H-NS proteins from the DNA-stiffening to the DNA-bridging mode, which has been observed repeatedly upon addition of multivalent cations to the buffer, but is still not understood. Unravelling the corresponding mechanism is all the more crucial, as the regulation properties of H-NS proteins, as well as other nucleoid proteins, are linked to their DNA-binding properties. The simulations reported here support a mechanism, according to which the primary role of multivalent cations consists in decreasing the strength of H-NS/DNA interactions compared to H-NS/H-NS interactions, with the latter ones becoming energetically favored with respect to the former ones above a certain threshold of the effective valency of the cations of the buffer. Below the threshold, H-NS dimers form filaments, which stretch along the DNA molecule but are quite inefficient in bridging genomically distant DNA sites (DNA-stiffening mode). In contrast, just above the threshold, H-NS dimers form 3-dimensional clusters, which are able to connect DNA sites that are distant from the genomic point of view (DNA-bridging mode). The model provides clear rationales for the experimental observations that the switch between the two modes is a threshold effect and that the ability of H-NS dimers to form higher order oligomers is crucial for their bridging capabilities.

[(#)] marc.joyeux@univ-grenoble-alpes.fr

# INTRODUCTION

Nucleoid Associated Proteins (NAP) are DNA-binding proteins with low to medium sequence specificity, which participate actively in bacteria' chromosome organization and gene expression regulation by bridging, bending, or caging the DNA molecule (1-3). Despite continuous efforts of several groups (see (1-3) and references therein), the mechanisms by which NAP achieve these goals remain in most cases elusive and controversial. Part of the difficulty arises from the fact that these mechanisms appear to be specific to each NAP. For example, the histone-like nucleoid structuring protein (H-NS), ParB, and structural maintenance of chromosome proteins (SMC) are all believed to bridge DNA, but the bridges are qualitatively different, which results in different effects on chromosome organization and gene expression regulation (4). Even worse, a single NAP may display several binding modes to DNA, depending on a variety of external factors, like pH, temperature, and the composition of the buffer (5-9). For example, atomic force microscopy and optical tweezers experiments first suggested that H-NS binding to DNA leads to the formation of bridges between DNA duplexes (this is the so-called *bridging* mode) (5), while subsequent magnetic tweezers experiments instead concluded that the DNA molecule rather adopts a more extended and stiffer conformation upon binding of the proteins (this is the so-called *stiffening* mode) (6). It was later pointed out that the discrepancy between the two sets of experiments may arise from the fact that the buffer used in the first set of experiments contained divalent salt cations, while the buffer used in the second set did not, and it was accordingly shown that both binding modes do exist and that a switch between the two of them can be driven by changes in divalent cations concentrations (7). More recent small angle neutron scattering experiments (8), as well as experiments performed in confined geometries (9), confirmed the crucial role of divalent cations. Still, it should be pointed out, as stated in (7), that "*the specific mechanism by which magnesium and calcium ions alter H-NS binding properties is currently unknown*", which is an all the more regrettable lack, as the regulation properties of H-NS (10,11), as well as other NAP (12,13), are probably linked to their DNA-binding properties. The purpose of the present paper is to propose an explanation for the role of divalent cations, based on our current knowledge of the properties of H-NS proteins and the results of simulations performed with a coarse-grained model, which was developed specifically for this purpose.

H-NS is a small protein (137 residues, 15.5 kDa), which is functional as a dimer. Each monomer consists of a N-terminal oligomerization domain (residues 1-64) (14,15) and a C-terminal DNA-binding domain (residues 91-137) (16) connected by a flexible linker (17). A secondary dimerization site at the C-terminal end of the main oligomerization domain allows H-NS dimers to organize in superhelical chains in crystals (18) and is probably also responsible for the oligomerization of H-NS in solution, where dimers, tetramers, and larger oligomers have been observed under different conditions (14,19,20). A thermodynamic analysis of such solutions led to a value of the enthalpy change upon dimerization or tetramerization of H-NS proteins of the order of $10\,k_B T$ at room temperature (20), which is very close to the value that was reported for the enthalpy change upon formation of a complex between DNA and a H-NS protein in solution (21). As will be developed below, this similarity of the values of the enthalpy changes upon oligomerization of H-NS and binding of H-NS to DNA is crucial for the dynamics of H-NS/DNA mixtures.

A second important point deals with the very peculiar properties of DNA when immersed in a solution containing dilute cations. Naked DNA is indeed a highly charged polyanion with two phosphate groups per base pair, resulting in a bare linear charge density of about $-5.9\,\bar{e}/\text{nm}$, where $\bar{e}$ is the absolute charge of the electron. However, the Manning-Oosawa condensation theory (22,23) stipulates that the net linear charge density along a polyion immersed in a buffer containing counterions of valency $Z$ cannot be larger than $\bar{e}/(Z\ell_B)$, where $\ell_B = \bar{e}^2/(4\pi\varepsilon_0\varepsilon_r k_B T)$ denotes the Bjerrum length of the buffer, that is, the distance at which the electrostatic interaction between two elementary charges $\bar{e}$ is equal to the thermal energy $k_B T$. If the bare charge density along the naked polyion is larger than the critical value $\bar{e}/(Z\ell_B)$, then an instability occurs and counterions coalesce on the polyion and neutralize an increasing number of its charges, till the net density reduces to the critical value $\bar{e}/(Z\ell_B)$. Owing to its relative dielectric constant $\varepsilon_r \approx 80$, the Bjerrum length of water at 25°C is $\ell_B \approx 0.7$ nm and the critical charge density $\approx -1.4\,\bar{e}/\text{nm}$ (for a buffer with monovalent cations) or $\approx -0.7\,\bar{e}/\text{nm}$ (for a buffer with divalent cations). This implies that (i) counterions do coalesce on the DNA molecule for both monovalent and divalent cations, and (ii) increasing the valency of the cations leads to a proportional reduction of the net charge density along the DNA.

Counterion condensation is expected to have important consequences on H-NS/DNA interactions. Indeed, except in the vicinity of a few high-affinity binding sites (24), H-NS/DNA interactions are mainly non-specific, as indicated by the very small variation of the change in heat capacity with temperature in the range 10-25°C (21) and by the fact that protein occupation on DNA decreases as monovalent salt concentration increases (7). H-NS proteins and cationic counterions therefore compete for binding to the DNA and theory indicates that the binding of polypeptides to the DNA is indeed accompanied by the release of counterions in the buffer (25-28). The counterions regain translational entropy upon release from the DNA (25-28), so that the net energy balance for the binding of ligands to the DNA results from subtle enthalpy-entropy compensations (29,30). Of outmost importance for the present work is the fact that this balance depends sensitively on the valency of the cationic counterions present in the buffer. It has indeed repeatedly been observed that addition of small amounts of divalent cations provokes a substantial decrease of the free energy of binding of polypeptides and proteins to nucleic acids (31-33).

In contrast, the bare charge density is much smaller along proteins than along nucleic acids, usually smaller than Manning-Oosawa's critical density, so that counterion condensation is expected to be much less marked for proteins than for nucleic acids, although there are theoretical indications that it might not totally vanish (34). Still, the short length of H-NS proteins should contribute to further diminish the eventual importance of counterion condensation (34), if it does take place. As a consequence, H-NS/H-NS interactions are expected to remain mostly insensitive to the valency of the cations of the buffer. More precisely, both H-NS/H-NS and H-NS/DNA interactions depend on the ionic strength of the buffer through the variation of the Debye length, but H-NS/DNA interactions depend additionally very sensitively on the valency of the cations through the mechanisms of counterion condensation and release.

Considered together, the arguments sketched above consequently suggest that binding of H-NS to DNA is favored in buffers with monovalent cations, while oligomerization of H-NS proteins is favored in buffers containing substantial amounts of divalent cations, *the role of divalent cations (like magnesium and calcium ions) consisting precisely in displacing the equilibrium from predominant H-NS/DNA interactions towards predominant H-NS/H-NS interactions*.

While this assumption is the key point of the present work, there still remains to understand why displacing the equilibrium towards leading H-NS/H-NS interactions drives H-NS from the stiffening to the bridging mode. In the same spirit as for previous work dealing with facilitated diffusion (35-37) and the compaction of genomic DNA inside the nucleoid (38-42), a mesoscopic beads and springs-type model was developed specifically to answer this question. This model incorporates the effective valency of the cations, $Z$, as a free parameter and simulations were launched with different values of $Z$. The results presented in Results and Discussion highlight the fact that a small variation of $Z$ is indeed able to induce strong changes in the conformations of the system. More precisely, in the range of values of *Z* where H-NS/DNA interactions prevail over H-NS/H-NS interactions, the dynamics of the system is driven by the attachment of protein chains to the DNA chain in *cis* configuration, their sliding along the DNA, and the formation of filaments of proteins along the DNA duplex, with the length and internal connectivity of the filaments increasing with *Z*. Such filaments have repeatedly been reported as a characteristic of the stiffening mode (6,7,10). For slightly larger values of *Z*, H-NS oligomerization however takes over the binding of H-NS to DNA and the system rather organizes in the form of 3-dimensional H-NS clusters, which efficiently bind regions of the DNA duplex that are broadly separated from the genomic point of view. This is the bridging mode, which leads to the collapse of the DNA for sufficiently large protein concentrations.

This work consequently supports a mechanism for the switch of H-NS proteins from the DNA-stiffening mode to the DNA-bridging mode upon increase of the concentration of divalent cations, which consists of (i) the displacement of the equilibrium from predominant H-NS/DNA interactions towards predominant H-NS/H-NS interactions, and (ii) the resulting switch of the preferred organization of H-NS proteins from filaments stretching along a genomic contiguous part of the DNA molecule to clusters able to connect parts of the DNA molecule that are genomically broadly separated.

**METHODS**

The mesoscopic model which has been developed for the present study is described in detail in Model and Simulations in the Supporting Material. In brief, the DNA is modelled as a circular chain of 2880 beads with radius 1.0 nm separated at equilibrium by a distance 2.5 nm and enclosed in a sphere with radius 120 nm. Two beads represent 15 DNA base pairs. Both the contour length of the DNA molecule and the cell volume are reduced by a factor of approximately 200 with respect to their actual values in *E. coli* cells, so that the nucleic acid concentration of the model is close to the *in vivo* one. DNA beads interact through stretching, bending, and electrostatic terms. The bending rigidity constant is chosen so that the model reproduces the known persistence length of DNA (50 nm). Electrostatic repulsion between DNA beads is written as a sum of Debye-Hückel terms, which depend on effective electrostatic charges placed at the center of each bead. These charges are assumed to be inversely proportional to a parameter *Z*, which represents the effective valency of the cations of the buffer, in order to account for counterion condensation along the DNA molecule. *Z* is equal to 1 (respectively, 2) for monovalent (respectively, divalent) cations, but may take any real value between 1 and 2 when divalent cations are added to a buffer that contains monovalent cations. This point is discussed in more detail in Results and Discussion. Repulsion between DNA beads decreases as $1/Z^2$ upon increase of $Z$.

H-NS dimers are modeled as chains of 4 beads with radius 1.0 nm separated at equilibrium by a distance 4.0 nm. For most simulations, 200 protein chains were introduced in the confining sphere together with the DNA chain, which corresponds to a protein concentration approximately twice the concentration of H-NS dimers during the cell growth phase and six times the concentration during the stationary phase (43). Protein chains have internal stretching and bending energy and interact with each other and with the DNA chain. The value of the bending constant is assumed to be as low as $2\,k_\text{B}T$, in order to account for the flexible linker that connects the C-terminal and N-terminal domains of H-NS (17).

A major approximation of the model consists in assuming that the interactions among protein beads and between protein beads and DNA beads are mediated uniquely by effective electrostatic charges placed at the center of each protein bead and to disregard all interactions beyond the crude electrostatic ones. The charges are positive for the two terminal beads of each chain and negative for the two central beads. The terminal beads of each protein chain can therefore bind either to the beads of the DNA chain or to the central beads of other protein chains, so that the model accounts for both H-NS oligomerization and binding of H-NS to DNA. The values of the charges at the center of protein beads are assumed to be independent of $Z$, thereby reflecting the fact that counterion condensation on proteins is negligible. Consequently, the attraction term between terminal beads of a protein chain and central beads of another protein chain does not depend on $Z$, while the attraction term between DNA beads and terminal protein beads evolves as $1/Z$. This is one of the key points of the model.

A Lennard-Jones-type excluded volume term is added to the attractive Debye-Hückel term for pairs of beads with opposite charges. For the sake of simplicity, this excluded volume potential is assumed to be independent of $Z$ and identical for protein/protein and protein/DNA pairs of beads. The two parameters of the potential were adjusted manually in order that the enthalpy changes upon forming a complex between two protein chains and between a protein chain and the DNA chain are comparable to the experimentally determined value for H-NS (20,21). As shown in Figs. S1 and S2 of the Supplemental Material, this enthalpy change is equal to $-12.0\,k_\text{B}T$ both for two protein chains approaching one another perpendicularly, and for a protein chain approaching the linear DNA chain perpendicularly at $Z=1.37$. Since H-NS/DNA attraction decreases like $1/Z$, while H-NS/H-NS attraction does not depend on $Z$, it is expected that this value $Z = 1.37$ plays a critical role in the model, with the formation of H-NS/DNA bonds being energetically favored for $Z < 1.37$ and oligomerization of H-NS being favored for $Z > 1.37$. This particular value $Z = 1.37$ will therefore be labelled $Z_{\text{crit}}$ in the remainder of the paper. The evolution with $Z$ of the enthalpy change upon formation of a H-NS/DNA bond is shown as a solid line in Fig. 1 and the enthalpy change upon formation of a H-NS/H-NS bond as a horizontal dot-dashed line. The dashed line indicates the energy of the saddles that separate two minima, see Fig. S2. The distance between the solid and dashed lines therefore represents the gap that proteins must overcome to translocate from one DNA binding site to the next one.

It is worth emphasizing that the model proposed here relies on a rather crude approximation, in the sense that it ignores most of the complexity of the binding of H-NS proteins to the DNA. The model indeed assumes that binding dynamics is governed by electrostatic interactions between protein beads and the DNA chain, which carries the critical charge density for a given value of $Z$, and disregards more complex mechanisms, like the release of counterions. This model was used in spite of its naivety, because it nonetheless reproduces the marked decrease of the H-NS/DNA binding energy upon addition of multivalent cations, which is central to this work.

The dynamics of the complete system was investigated by integrating numerically the Langevin equations of motion with kinetic energy terms dropped and with time step set to 1.0 ps for simulations with 200 protein chains and 0.5 ps for 1000 protein chains. After each integration step, the position of the centre of the confining sphere was slightly adjusted so as to coincide with the centre of mass of the DNA molecule. Temperature $T$ was assumed to be 298 K throughout the study. Representative snapshots for 200 protein chains and $Z = 1.33$, 1.42, and 1.50, are shown in Fig. 2.

**RESULTS AND DISCUSSION**

As mentioned in the Introduction, the Manning-Oosawa condensation theory (22,23) stipulates that the effective linear charge along a highly charged polyion immersed in a buffer containing counterions of valency $Z$ is $\bar{e}/(Z\ell_B)$ as soon as counterion condensation occurs. The case of a buffer containing cations of two different valencies $Z_1$ and $Z_2$ is more complex, because the cations compete for condensation. Still, the fractions of DNA phosphate sites neutralized by cations of type 1 and 2 ($\theta_1$ and $\theta_2$, respectively) can be readily obtained from the equations derived in (44). One can then compute the effective valency of the cations of the buffer, according to

$$Z = \frac{b}{\ell_B (1 - Z_1\theta_1 - Z_2\theta_2)} , \qquad (1)$$

where $b = 0.17$ nm is the effective distance between two charges along naked DNA, and $Z_1\theta_1 + Z_2\theta_2$ the total fraction of DNA phosphate charges neutralized by the different cations. $Z$ is always comprised between $Z_1$ and $Z_2$ and the net linear charge density along the DNA backbone is $-\overline{e}/(Z\ell_B)$. Application of Eq. (1) to a buffer containing 10mM of $MgCl_2$ in addition to 60mM of KCl, as was used in (5) and also investigated in (31-33), leads to an effective valency $Z = 1.63$, meaning that addition of a relatively small amount of divalent cations results in a significant increase of $Z$. Addition of this amount of $MgCl_2$ also increases the Debye length by about 30%, but this variation affects H-NS/DNA and H-NS/H-NS interactions essentially in the same way, so that this effect will be disregarded in the present simulations, which focus on the effects of varying $Z$ on the dynamics of the system.

In the model, the effective valency $Z$ is actually considered as a free parameter, which affects only the charge placed at the center of each DNA bead, this charge being obtained as the product of the equilibrium distance between two beads and the net linear charge density $-\overline{e}/(Z\ell_B)$. Simulations were run for 9 different values of $Z$ ranging from 1.0 to 1.67 and two different numbers of protein chains (200 and 1000), while keeping all other parameters of the system constant. Representative snapshots of the conformations obtained with 200 proteins chains and $Z$=1.33, 1.42, and 1.50, are shown in Fig. 2. It is reminded that H-NS/DNA interactions are energetically favored for $Z < Z_{crit} = 1.37$, while H-NS/H-NS interactions are favored for $Z > Z_{crit}$. Visually, one indeed observes a qualitative difference between the three snapshots in Fig. 2, with the protein chains organizing principally in short filaments stretching along the DNA for $Z$=1.33, in rather loose clusters connected to the DNA for $Z$=1.42, and in bigger, more compact clusters connected to the DNA for $Z$=1.50.

A more quantitative insight into the evolution of the conformations of the system can be gained by plotting, for increasing values of $Z$, the probability distribution $p(s)$ for a protein chain to be bound to $s$ DNA beads (Fig. 3, left column), the probability distribution $q(s)$ for a protein chain to be bound to $s$ other protein chains (Fig. 3, right column), as well as the probability distribution $u(s)$ for this protein chain to belong to a cluster composed of $s$ protein chains (Fig. 4) (the closely related plot of the size distribution of protein clusters is shown is Fig. S3 of the Supporting Material). For this purpose, it was considered that the protein chain is bound to a DNA bead if their interaction is attractive and of total magnitude (computed according to Eq. (S11)) larger than $3\,k_\text{B}T$, and that two protein chains are bound if the interaction between one chain and one terminal bead of the other chain is attractive and of total magnitude (computed according to Eq. (S9)) larger than $3\,k_\text{B}T$. The choice of the threshold is somewhat arbitrary but the principal features of the distributions do not depend critically thereon. It is seen in Fig. 3 that for $Z$=1.00, that is, when H-NS/DNA interactions are energetically favored with respect to H-NS/H-NS ones, most protein chains bind to four DNA beads but do not bind to other protein chains. Remembering that tight binding of a protein chain to the DNA chain involves two consecutive DNA beads (see Fig. S2), this indicates that most protein chains bind the DNA chain with their two terminal beads (that is, in *cis*) while remaining separated from the other protein chains. This is confirmed by the distribution of cluster sizes (Fig. 4), which indicates that about 70% of the protein chains belong to clusters of size 1 (and about 25% to clusters of size 2). In contrast, for $Z$=1.67, that is, when H-NS/H-NS interactions are energetically favored with respect to H-NS/DNA ones, most protein chains do not bind to the DNA chain but bind instead on average to 5 other protein chains (Fig. 3). This suggests that protein chains form large clusters, which interact only loosely with the DNA chain. The distribution of cluster sizes (Fig. 4) accordingly indicates that large clusters with size up to 100 chains sequestrate the majority of protein chains.

For 200 protein chains, the transition between these two limiting regimes occurs around $Z_\text{crit}=1.37$, as may be checked in Fig. 3. In this range of values of $Z$, many protein chains bind to one other protein chain and form simultaneously one strong bond (involving two DNA beads) with the DNA chain. This suggests that filaments of connected protein chains form along the DNA chain, where each protein chain binds with one terminal bead to the DNA chain and with the other terminal bead to the neighboring protein chain. The plot of $u(s)$ in Fig. 4 accordingly indicates that, for $Z$=1.33, about 35% of the protein chains belong to filaments of size 3 to 6 (and about 35% to filaments of size 2). These filaments are admittedly not very long and they are furthermore dispersed rather randomly over the whole contour length of the DNA chain. It should however be reminded that 200 protein chains correspond approximately to one H-NS dimer per 100 DNA base pair, that is, about twice the physiological concentration during the cell growth phase (43). This concentration is much smaller than the typical protein concentrations used in (6,7,9,10), which were of the order of one H-NS dimer per one to ten DNA base pairs and led to the collapse of the DNA molecule when no stretching force was applied to its extremities. Simulations indicate that, for 1000 protein chains, filaments form at lower values of $Z$, are much longer, and cover significantly larger portions of the DNA contour length, as may be checked in Fig. 5, which shows a representative snapshot of the system for $Z$=1.17.

Careful analysis of the results of simulations indicate that the mechanical properties of the DNA chain (like its persistence length) are not significantly altered by the assembly of protein filaments on the DNA chain, even for 1000 protein chains. This is probably due to the fact that the bending rigidity of the protein chains was assumed to be much smaller than the rigidity of the DNA chain, in order to account for the flexible linker that connects the C-terminal and N-terminal domains of H-NS (17). As a consequence, the protein chains and the filaments merely adapt to the deformations of the DNA chain without hampering them significantly. This property of the model is in clear contradiction with experimental results, which indicate that the formation of filaments is accompanied by an increase in the persistence length and stiffness of the DNA molecule (6,7,10). Such a discrepancy between the results of simulations and experiments strongly suggests that protein filaments are in themselves not responsible for the increased stiffness and that the mechanism behind it is more probably related to the way H-NS dimers bind to the DNA duplex. More precisely, it has been shown that H-NS dimers insert one C-terminal loop inside the minor groove of double-stranded DNA (45). Binding of a H-NS dimer to a DNA site consequently decreases the flexibility of the DNA chain at this particular location, a point not accounted for in the model, and the many H-NS/DNA bonds associated with the formation of protein filaments are probably responsible for the observed increase in overall DNA stiffness.

Another important feature of protein filaments is that they stretch along the DNA chain but only seldom bridge sites that are broadly separated from the genomic point of view. However, the geometry of protein clusters changes drastically above $Z_{\text{crit}} = 1.37$, where H-NS/H-NS interactions become energetically favored with respect to H-NS/DNA ones. Indeed, above $Z_{\text{crit}}$ protein clusters grow in three dimensions in the buffer, instead of growing in one dimension along the DNA chain, which enables them to form simultaneous contacts with several sites along the DNA chain that are not contiguous from the genomic point of view, as may be checked in the middle and bottom vignettes of Fig. 2. Particularly striking is the plot, as a function of $Z$, of the number of indirect connections between pairs of DNA beads mediated by H-NS clusters, which is shown in Fig. 6. Two DNA beads are considered to be connected if they are linked by protein chains that form a continuous series of contacts of the form *d-p-p-p-... -p-p-p-d*, where *d* denotes a DNA bead and *p* a protein chain, and there is no limit on the number of intercalated protein chains *p*. The same threshold as above is used for defining a contact ($3\,k_\text{B}T$ for both DNA/H-NS and H-NS/H-NS contacts), and only pairs of DNA beads *d* separated along the DNA chain (genomic separation) by 50 or more other DNA beads are taken into account (the result is essentially independent of the exact value of the separation threshold, as soon as it is larger than the contour length of the short protein filaments observed for 200 protein chains). It is seen in Fig. 6, that the number of connections increases sharply just above $Z_\text{crit}$, which indicates that connection of distant DNA sites by H-NS proteins is efficient only when H-NS/H-NS interactions are favored with respect to H-NS/DNA ones and H-NS proteins are able to form clusters in three dimensions.

Let us mention for the sake of completeness that other quantities, like the coverage of the DNA chain by protein chains, may however display smoother variations around $Z_\text{crit}$ than the quantities discussed above, as can be checked in Fig. S4 of the Supporting Material. For the specific case of the coverage of the DNA chain by protein chains, the smoother behavior is due to the fact that the formation of both protein filaments and 3-dimensional clusters contribute to reduce the coverage of DNA (because of the overlap of protein chains in the former case and of protein chains not being bound to the DNA chain in the latter case), so that coverage of DNA decreases steadily and rather uniformly over the full range of variation of *Z*.

It is worth emphasizing that the present work consequently describes the DNA-bridging mode of H-NS proteins as consisting of protein clusters that bind broadly separated sites along the DNA chain, and not, as is often more or less implicitly assumed, as tracts of parallel proteins that bridge two parallel DNA duplexes. This latter description emerges quite naturally when looking at atomic force micrographs of DNA/protein complexes deposited on mica plates (5), but simulations have shown that the conformations of DNA/protein complexes are significantly different in bulk and in planar conditions and that the parallel protein bridges geometry arises from the rearrangement of the complexes after their deposition on the charged surface (38,39). Moreover, recent single-molecule experiments have reported the presence of clusters of H-NS proteins buried inside the nucleoid (46), while other experiments have shown that mutants of H-NS, which have disrupted dimer-dimer interactions and cannot form higher order oligomers, are also unable to form bridges between two DNA duplexes (47). The ability for H-NS dimers to clusterize therefore appears as a prerequisite for their ability to bridge distant DNA sites, which strongly supports the conclusions of the present work. Finally, it has been shown in the same work that the transition from no bridging to complete bridging is very abrupt upon increase of the concentration of magnesium ions (47), which is again in agreement with the threshold effect observed in Fig. 6.

**CONCLUSION**

The coarse-grained model developed in the present work fully supports the mechanism proposed in the Introduction, according to which the primary role of divalent cations in the switch of H-NS proteins from the DNA-stiffening mode to the DNA-bridging mode consists in decreasing the strength of H-NS/DNA interactions with respect to the strength of H-NS/H-NS interactions, with the latter ones becoming energetically favored above a certain threshold of the effective valency. The geometry of H-NS clusters changes drastically at this threshold. Below the threshold, H-NS dimers form filaments, which stretch along the DNA molecule but are quite inefficient in bridging DNA sites that are not contiguous from the genomic point of view. This is the DNA-stiffening mode. In contrast, above the threshold, H-NS dimers form 3-dimensional clusters, which are able to connect DNA sites that are distant from the genomic point of view. This is the DNA-bridging mode.

The model does not reproduce the increase in DNA stiffness induced by the assembly of protein filaments on the DNA molecule, because it does not account for the local increase of rigidity associated with the insertion of the C-terminal loop of H-NS proteins inside the minor groove of the DNA molecule. In contrast, it provides clear rationales for the experimental observations that the switch between the two modes is a threshold effect and that the ability of H-NS dimers to further oligomerize is crucial for their bridging capabilities.

This work suggests that the switch of H-NS proteins between the DNA-stiffening and DNA-binding modes can be explained on the basis of simple, general arguments. According to the recent study quoted above (47), it may instead result from a more involved mechanism, where divalent cations let H-NS dimers switch between a "closed" and an "open" geometry, which have different oligomerization and DNA-binding properties (47). While it is not clear, why such a mechanism would display a threshold, it can of course not be excluded that both mechanisms cooperate to let H-NS proteins switch between the two DNA-binding modes.

Last but not least, it appears that the concentration of divalent cations where the switch occurs falls within the physiological range, which may be considered as an indication that the mechanism described here may play an important role in the global regulation scheme of bacterial cells.

**SUPPORTING MATERIAL**

Model and Simulations section. Figures S1 to S4.

**SUPPORTING CITATIONS**

Reference (48) appears in the Supporting Material.

# FIGURE CAPTIONS

**Figure 1** : Plot, as a function of the effective valency $Z$ of the cations of the buffer, of the minimum of the potential energy felt by a linear protein chain approaching the linear DNA chain perpendicularly (solid red curve) and of the energy of the saddle separating two minima (dashed blue curve). The potential energy surface itself is shown in Fig. S2 of the Supporting Material for $Z = Z_{\text{crit}} = 1.37$. The dot-dashed green horizontal line represents the minimum of the potential energy felt by a linear protein chain approaching another linear protein chain perpendicularly. The potential energy surface itself is shown in Fig. S1 of the Supporting Material. The solid and dot-dashed lines intersect at $Z = Z_{\text{crit}} = 1.37$.

**Figure 2** : Representative snapshots of the system with 200 proteins chains for $Z$=1.33 (top), 1.42 (middle) and 1.50 (bottom). The blue line connects the centers of successive DNA beads. Red (respectively, green) spheres represent terminal (respectively, central) beads of protein chains. The confining sphere is not shown.

**Figure 3** : Probability distribution $p(s)$ for a protein chain to be bound to $s$ DNA beads (left column) and probability distribution $q(s)$ for a protein chain to be bound to $s$ other protein chains (right column), for values of Z increasing from 1.00 to 1.67. Note that binding of a protein chain to the DNA chain usually involves two successive DNA beads. Each plot was obtained from a single equilibrated simulation with 200 protein chains by averaging over a time interval of 0.1 ms.

**Figure 4** : Decimal logarithm of $u(s)$, the probability distribution for a protein chain to belong to a cluster containing $s$ protein chains, for values of $Z$ increasing from 1.00 to 1.67. Each plot was obtained from a single equilibrated simulation with 200 protein chains by averaging over a time interval of 0.1 ms.

**Figure 5** : Representative snapshots of the system with 1000 proteins chains for $Z$=1.17. The blue line connects the centers of successive DNA beads. Red (respectively, green) spheres represent terminal (respectively, central) beads of protein chains. The confining sphere is not shown.

**Figure 6** : Plot, as a function of Z, of the average number of connections between pairs of DNA beads mediated by protein chains, for 200 protein chains. Only pairs of beads separated along the chain by 50 (or more) other DNA beads are taken into account.

**Figure 1**

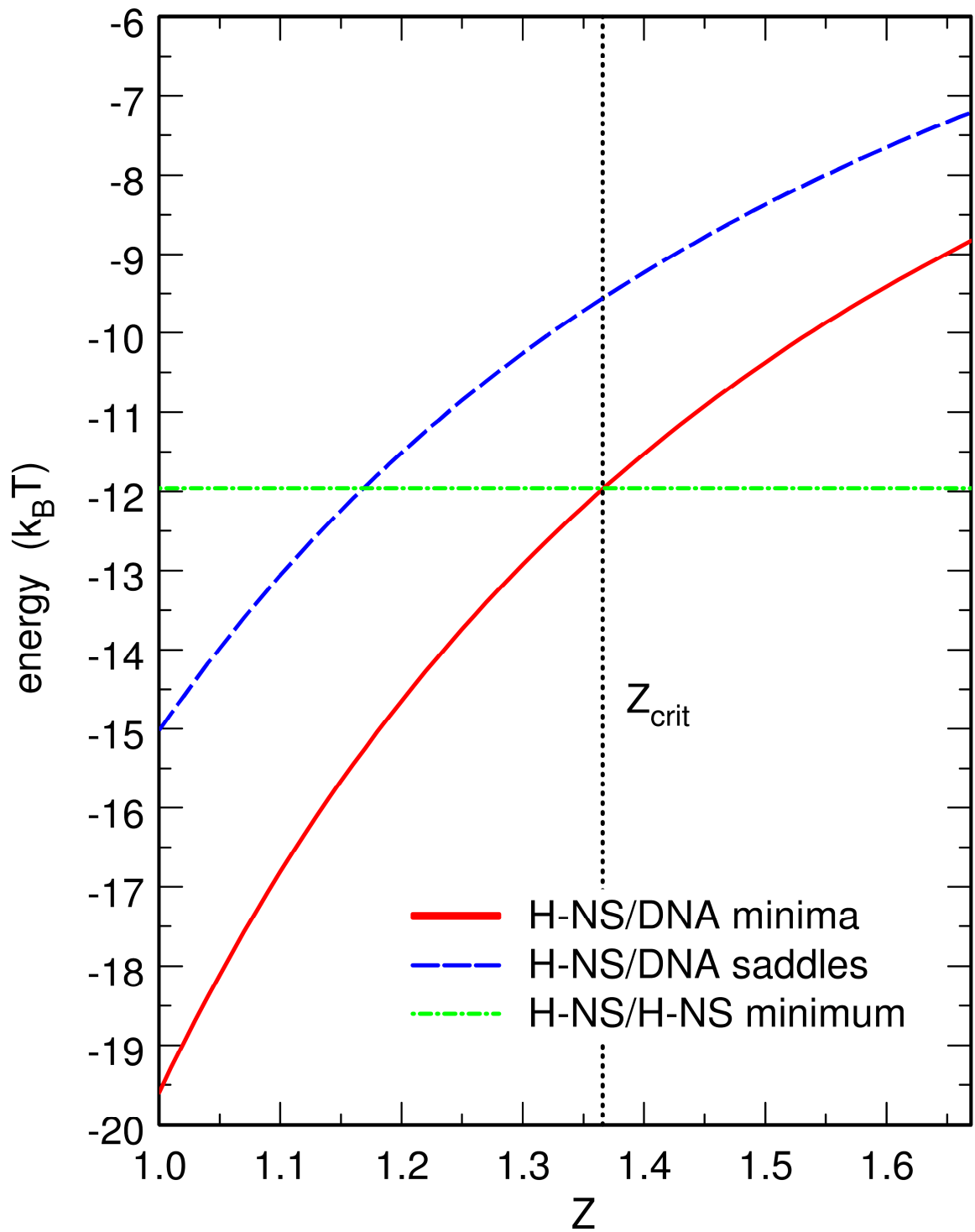

**Figure 2**

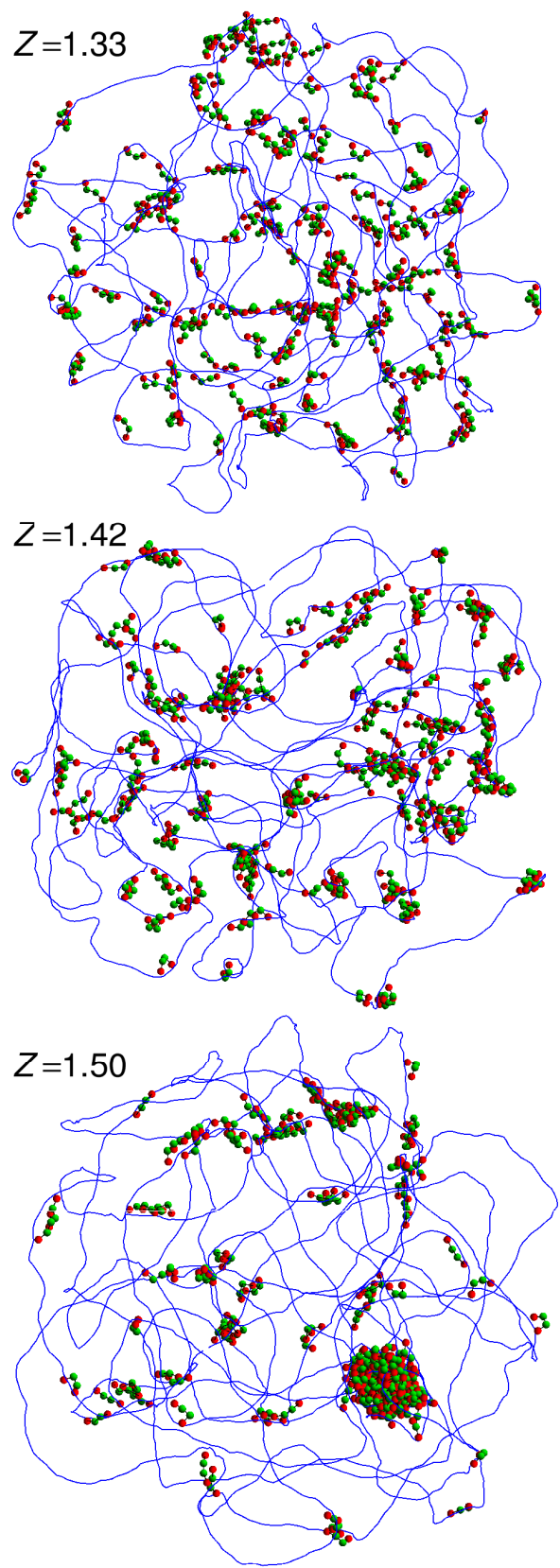

*Z* =1.33

*Z* =1.42

*Z* =1.50

**Figure 3**

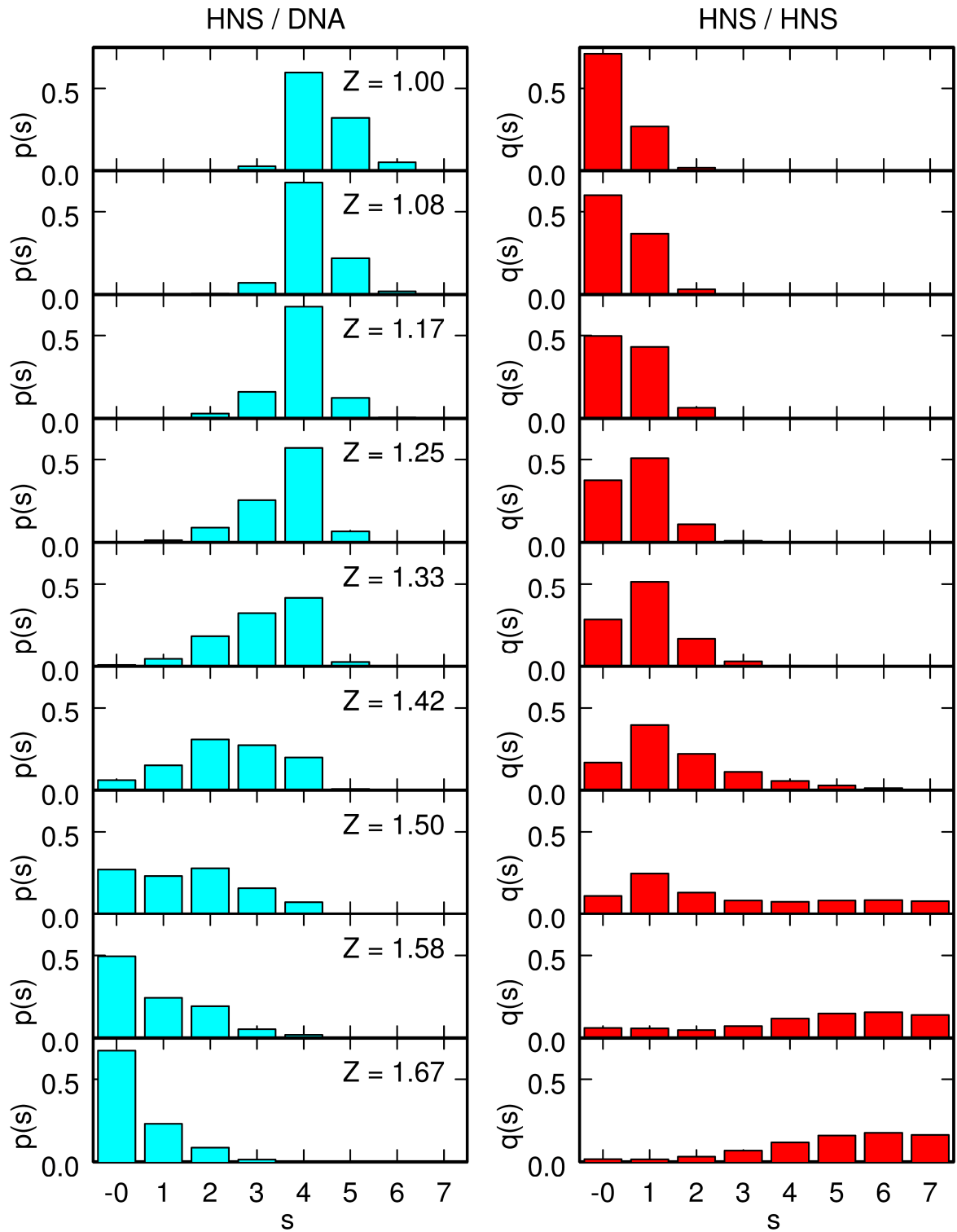

**Figure 4**

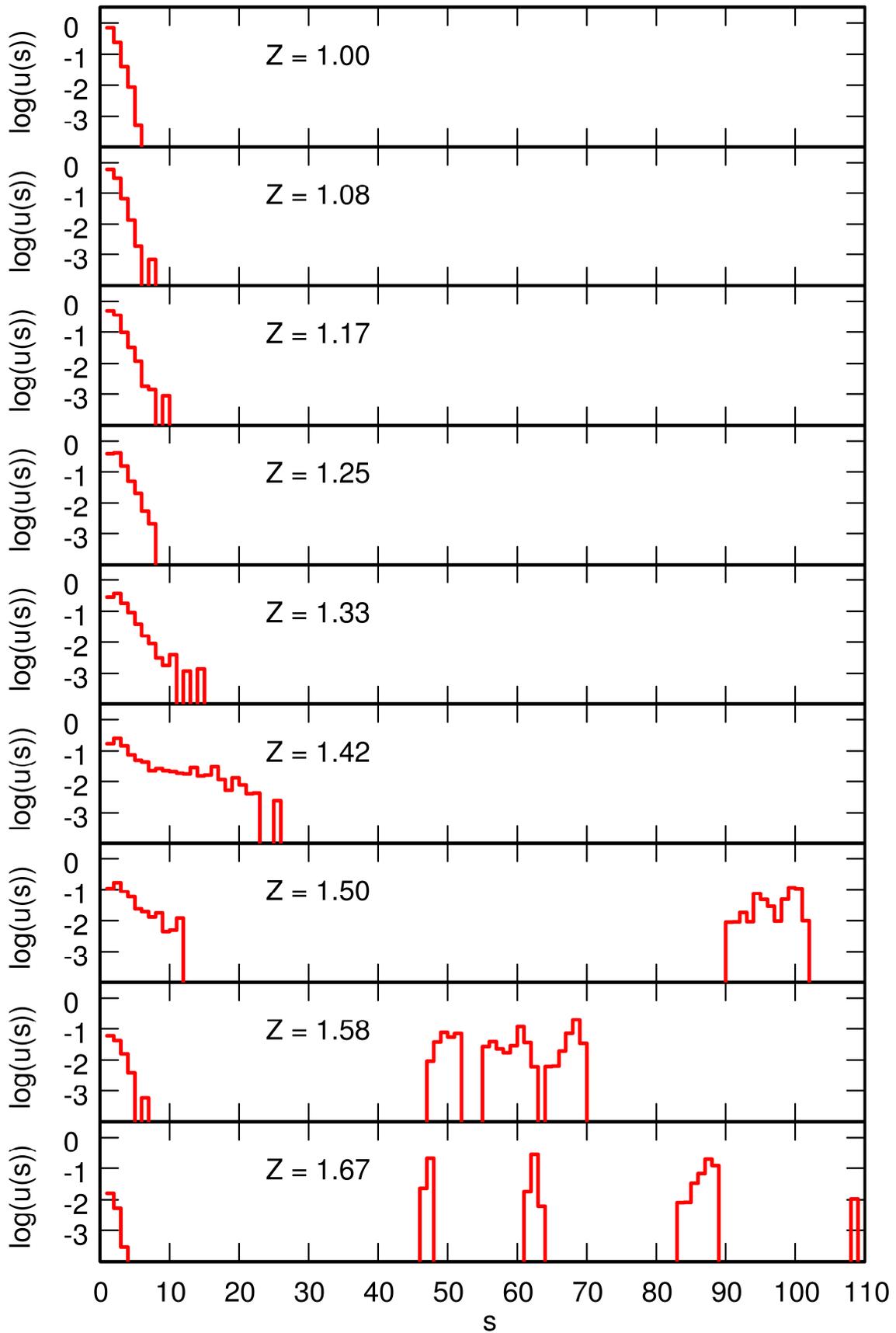

**Figure 5**

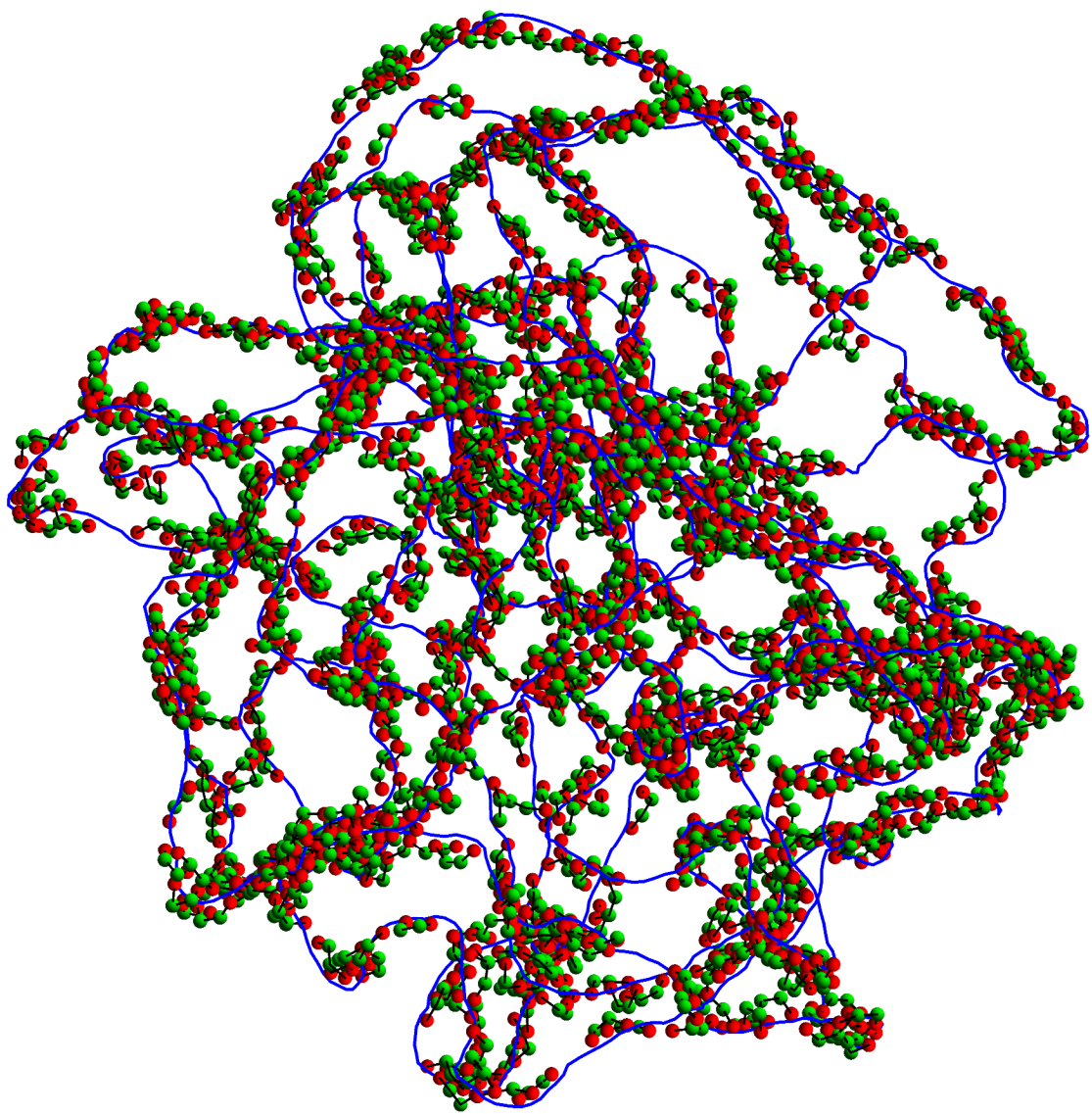

**Figure 6**

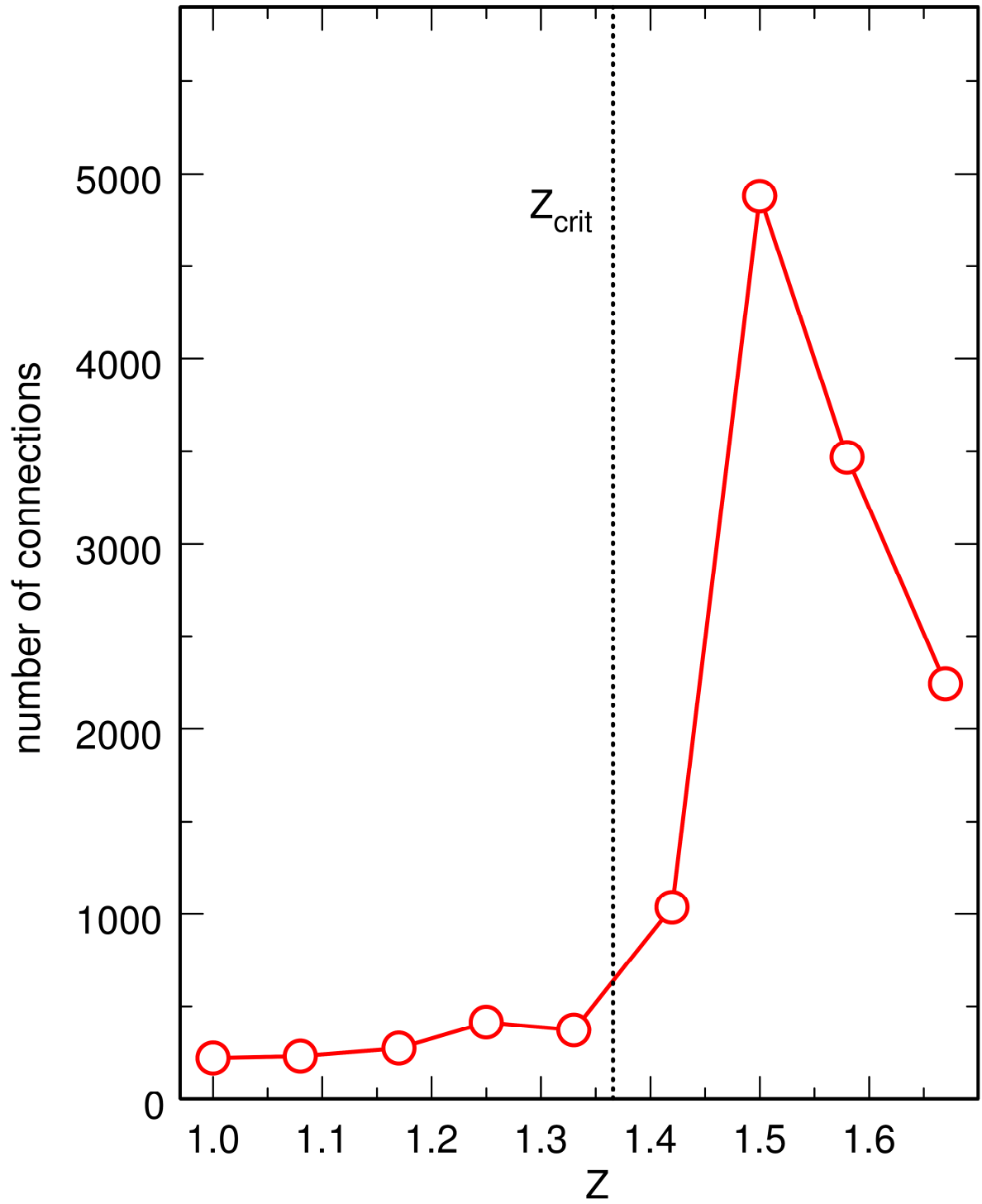

# Role of salt valency in the switch of H-NS proteins between DNA-bridging and DNA-stiffening modes

M. Joyeux

*LIPhy, CNRS and Université Grenoble Alpes, Grenoble, France*

## MODEL AND SIMULATIONS

Temperature $T$ is assumed to be 298 K throughout the study. The model for the DNA molecule consists of a circular chain of $n = 2880$ beads with radius $a = 1.0$ nm separated at equilibrium by a distance $l_0 = 2.5$ nm and enclosed in a sphere with radius $R_0 = 120$ nm. Two beads represent 15 DNA base pairs. The contour length of the DNA molecule and the cell volume correspond approximately to 1/200th of the values for *E. coli* cells, so that the nucleic acid concentration of the model is close to the physiological one. The potential energy of the DNA chain consists of 4 terms, namely, the stretching energy $V_s$, the bending energy $V_b$, the electrostatic repulsion $V_e$, and a confinement term $V_{wall}$

$$E_{DNA} = V_s + V_b + V_e + V_{wall} \ . \tag{S1}$$

The stretching and bending contributions write

$$V_s = \frac{h}{2} \sum_{k=1}^{n} (l_k - l_0)^2$$
$$V_b = \frac{g}{2} \sum_{k=1}^{n} \theta_k^2, \tag{S2}$$

where $\mathbf{r}_k$ denotes the position of DNA bead $k$, $l_k = \|\mathbf{r}_k - \mathbf{r}_{k+1}\|$ the distance between two successive beads, and $\theta_k = \arccos((\mathbf{r}_k - \mathbf{r}_{k+1})(\mathbf{r}_{k+1} - \mathbf{r}_{k+2})/(\|\mathbf{r}_k - \mathbf{r}_{k+1}\|\|\mathbf{r}_{k+1} - \mathbf{r}_{k+2}\|))$ the angle formed by three successive beads. The stretching energy $V_s$ is a computational device without biological meaning, which is aimed at avoiding a rigid rod description. The stretching force constant $h$ is set to $h = 100\, k_B T / l_0^2$, which insures that the variations of the distance between successive beads remain small enough (1). In contrast, the bending rigidity constant is obtained from the known persistence length of the DNA, $\xi = 50$ nm, according to $g = \xi k_B T / l_0 = 20 k_B T$.

Electrostatic repulsion between DNA beads that are not close neighbours along the chain is written as a sum of repulsive Debye-Hückel terms with hard core

$$V_e = (\frac{e_{DNA}}{Z})^2 \sum_{k=1}^{n-4} \sum_{K=k+4}^{n} H(\|\mathbf{r}_k - \mathbf{r}_K\|) ,\qquad(S3)$$

where

$$H(r) = \frac{1}{4\pi\varepsilon r} \exp\left(-\frac{r-2a}{r_D}\right) .\qquad(S4)$$

Interactions between close neighbours ($1 \leq |k - K| \leq 3$) are not included in Eq. (S3) because it is considered that they are already accounted for in the stretching and bending terms. $\varepsilon = 80\,\varepsilon_0$ denotes the dielectric constant of the buffer. The value of the Debye length $r_D$ is set to 3.07 nm, which corresponds to a concentration of monovalent salt of 0.01 M. $e_{DNA}$ is the electric charge, which is placed at the centre of each DNA bead when considering that the buffer contains only monovalent cations ($Z=1$). The numerical value $e_{DNA} = -3.525\,\bar{e}$, where $\bar{e}$ is the absolute charge of the electron, is the product of $l_0$ and the net linear charge density derived from Manning's counterion condensation theory ($-\bar{e}/\ell_B \approx -1.41\,\bar{e}/\text{nm}$, see the main text). The charge placed at the center of each bead reduces to $e_{DNA}/Z$ when considering that the buffer contains cations of effective valency $Z>1$.

Finally, the confinement term $V_{\text{wall}}$ is taken as a sum of repulsive terms

$$V_{\text{wall}} = 10\,k_B T \sum_{k=1}^{n} f(\|\mathbf{r}_k\|) ,\qquad(S5)$$

where $f$ is the function defined according to

if $r \leq R_0$ : $f(r) = 0$

if $r > R_0$ : $f(r) = \left(\frac{r}{R_0}\right)^6 - 1$ . (S6)

H-NS dimers are modeled as chains of 4 beads with radius $a$ separated at equilibrium by a distance $L_0 = 4.0$ nm. For each protein chain $j$, charges $e_{j1} = e_{j4} = 3\bar{e}$ are placed at the centre of terminal beads $m=1$ and $m=4$, and charges $e_{j2} = e_{j3} = -3\bar{e}$ at the centre of central beads $m=2$ and $m=3$. The values of these effective charges were obtained from a naive counting of the number of positively and negatively charged residues in published crystallographic structures (2). It is considered that the density of charges along the naked protein chain is small enough for counterion condensation not to take place in the range $1 \leq Z \leq 2$. Moreover, the terminal beads of each protein chain can bind either to the beads of the DNA chain or to the central beads of other protein chains, so that the model accounts for both H-NS oligomerization and binding of H-NS to the DNA chain. In most simulations, $P = 200$ protein chains were introduced in the confining sphere together with the DNA chain, which corresponds to a protein concentration approximately twice the concentration of H-NS dimers during the cell growth phase and six times the concentration during the stationary phase (3).

The potential energy of the protein chains consists of 4 terms

$$E_\mathrm{P} = V_s^{(\mathrm{P})} + V_b^{(\mathrm{P})} + V_e^{(\mathrm{P/P})} + V_\mathrm{wall}^{(\mathrm{P})}, \tag{S7}$$

where the stretching, bending, and confining energies are very similar to their DNA counterparts

$$\begin{aligned} V_s^{(\mathrm{P})} &= \frac{h}{2} \sum_{j=1}^{P} \sum_{m=1}^{3} (L_{jm} - L_0)^2 \\ V_b^{(\mathrm{P})} &= \frac{G}{2} \sum_{j=1}^{P} \sum_{m=1}^{m=2} \Theta_{jm}^2 \\ V_\mathrm{wall}^{(\mathrm{P})} &= 10\, k_\mathrm{B} T \sum_{j=1}^{P} \sum_{m=1}^{4} f(\|\mathbf{R}_{jm}\|), \end{aligned} \tag{S8}$$

with $\mathbf{R}_{jm}$ the position of bead $m$ of protein chain $j$, $L_{jm}$ the distance between beads $m$ and $m+1$ of protein chain $j$, and $\Theta_{jm}$ the angle formed by beads $m$, $m+1$, and $m+2$ of protein chain $j$. The value of the bending constant is assumed to be as low as $G = 2\, k_\mathrm{B} T$, in order to account for the flexible linker that connects the C-terminal and N-terminal domains of H-NS.

The interaction energy between protein chains, $V_e^{(\mathrm{P/P})}$, is taken as the sum of (attractive or repulsive) Debye-Hückel terms with hard core and (repulsive) excluded volume terms, with the latter ones contributing only if the corresponding Debye-Hückel term is attractive

$$\begin{aligned} V_e^{(\mathrm{P/P})} &= \sum_{j=1}^{P} e_{j1} e_{j4} H(\|\mathbf{R}_{j1} - \mathbf{R}_{j4}\|) + \sum_{j=1}^{P-1} \sum_{m=1}^{4} \sum_{J=j+1}^{P} \sum_{M=1}^{4} e_{jm} e_{JM} H(\|\mathbf{R}_{jm} - \mathbf{R}_{JM}\|) \\ &+ \chi \sum_{j=1}^{P} \sum_{m=1,4} \sum_{\substack{J=1 \\ J \neq j}}^{P} \sum_{M=2,3} F(\|\mathbf{R}_{jm} - \mathbf{R}_{JM}\|), \end{aligned} \tag{S9}$$

where $F$ is the function defined according to

if $r \leq r_0$ : $\quad F(r) = \dfrac{r_0 - 2a}{r - 2a}(\dfrac{r_0 - 2a}{r - 2a} - 2) + 1 \tag{S10}$

if $r > r_0$ :    $F(r) = 0$ ,

and $r_0$ denotes the threshold distance below which the excluded volume term, taken as the repulsive part of a $2^h$ order Lennard-Jones-like function with hard core, creates a repulsion force between oppositely charged beads. The first term in the right-hand side of Eq. (S9) insures that the two terminal beads of the same chain do not overlap. The numerical values of the two parameters of the excluded volume potential, $\chi = 1\, k_B T$ and $r_0 = 3.5$ nm, were adjusted manually in order that the enthalpy change upon forming a complex between two protein chains is comparable to the experimentally determined value for H-NS. As shown in Fig. S1, this enthalpy change is equal to $-12.0\, k_B T$ for two protein chains at equilibrium approaching one another perpendicularly, which is close to the experimentally determined value for H-NS ($\approx -10.2\, k_B T$ (4)).

Finally, the potential energy describing the interactions between the DNA chain and the protein chains, $E_{\text{DNA/P}}$, is similarly taken as the sum of (attractive or repulsive) Debye-Hückel terms with hard core and (repulsive) excluded volume terms, with the latter ones contributing only if the corresponding Debye-Hückel term is attractive

$$E_{\text{DNA/P}} = \frac{e_{\text{DNA}}}{Z} \sum_{k=1}^{n} \sum_{j=1}^{P} \sum_{m=1}^{4} e_{jm} H(\|\mathbf{r}_k - \mathbf{R}_{jm}\|) + \chi \sum_{k=1}^{n} \sum_{j=1}^{P} \sum_{m=1,4} F(\|\mathbf{r}_k - \mathbf{R}_{jm}\|) . \quad \text{(S11)}$$

It is emphasized that the attraction term between DNA beads and terminal protein beads scales as $1/Z$ in Eq. (S11), while the attraction term between terminal beads of a protein chain and central beads of another protein chain does not depend on $Z$ in Eq. (S9). This is a key point of the model, see the main text. The potential energy felt by protein chains at equilibrium approaching perpendicularly the linear DNA chain at equilibrium is shown in Fig. S2 for an effective valency of the cations $Z = 1.37$. For this particular value of $Z$, the enthalpy change upon binding of a protein chain to the DNA chain is $-12.0\, k_B T$, which is equal to the enthalpy change upon binding of a protein chain to another protein chain, and comparable to experimentally determined values ($\approx -11.0\, k_B T$ (5)). In contrast, as illustrated in Fig. 1 of the main text, binding of protein chains to the DNA chain is favored for $Z < 1.37$, while binding to other protein chains is favored for $Z > 1.37$. It may also be noted in this figure that the evolution of the enthalpy change upon binding of H-NS to DNA deviates slightly from an $1/Z$ law, which results from the fact that the excluded volume term is assumed to be independent of Z in Eq. (S11).

The total potential energy of the system, $E_{pot}$, is the sum of the energies of DNA and protein chains and DNA/protein interactions

$$E_{pot} = E_{DNA} + E_P + E_{DNA/P} \ . \tag{S12}$$

The dynamics of the model was investigated by integrating numerically the Langevin equations of motion with kinetic energy terms neglected. Practically, the updated position vector for each bead (whether DNA or protein), $\mathbf{r}_j^{(n+1)}$, is computed from the current position vector, $\mathbf{r}_j^{(n)}$, according to

$$\mathbf{r}_j^{(n+1)} = \mathbf{r}_j^{(n)} + \frac{D_t \, \Delta t}{k_B T} \mathbf{F}_j^{(n)} + \sqrt{2 D_t \, \Delta t} \, \xi^{(n)} \ , \tag{S13}$$

where the translational diffusion coefficient $D_t$ is equal to $(k_B T)/(6\pi\eta a)$ and $\eta = 0.00089$ Pa s is the viscosity of the buffer at $T = 298$ K. $\mathbf{F}_j^{(n)}$ is the vector of inter-particle forces arising from the potential energy $E_{pot}$, $\xi^{(n)}$ a vector of random numbers extracted at each step $n$ from a Gaussian distribution of mean 0 and variance 1, and $\Delta t$ the integration time step, which is set to 1.0 ps for $P = 200$ protein chains and 0.5 ps for $P = 1000$ chains. After each integration step, the position of the center of the confining sphere was slightly adjusted so as to coincide with the center of mass of the DNA molecule.

# SUPPORTING REFERENCES

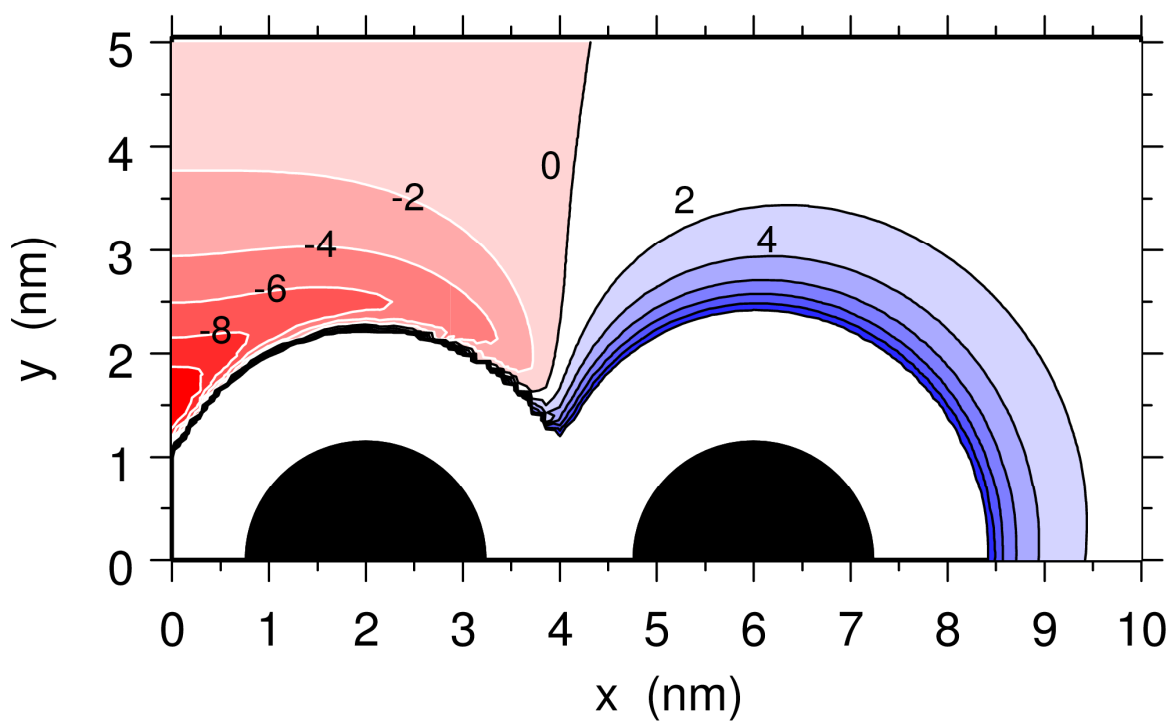

**Figure S1** : Potential energy felt by a protein chain aligned along the *y* axis when approaching another protein chain elongated along the *x* axis and centered on (0,0). Both chains are at equilibrium with respect to their stretching and bending degrees of freedom. The black disks represent two beads of the protein chain elongated along the *x* axis. In this geometry, the potential is symmetric with respect to the *y* axis, in addition to having rotational symmetry along the *x* axis. (*x*,*y*) denote the coordinates of the center of the terminal bead of the vertical chain that lies closest to the horizontal chain. The minimum of the potential energy surface ($-12.0\, k_B T$) is located on the *y* axis. Contour lines are separated by $2\, k_B T$.

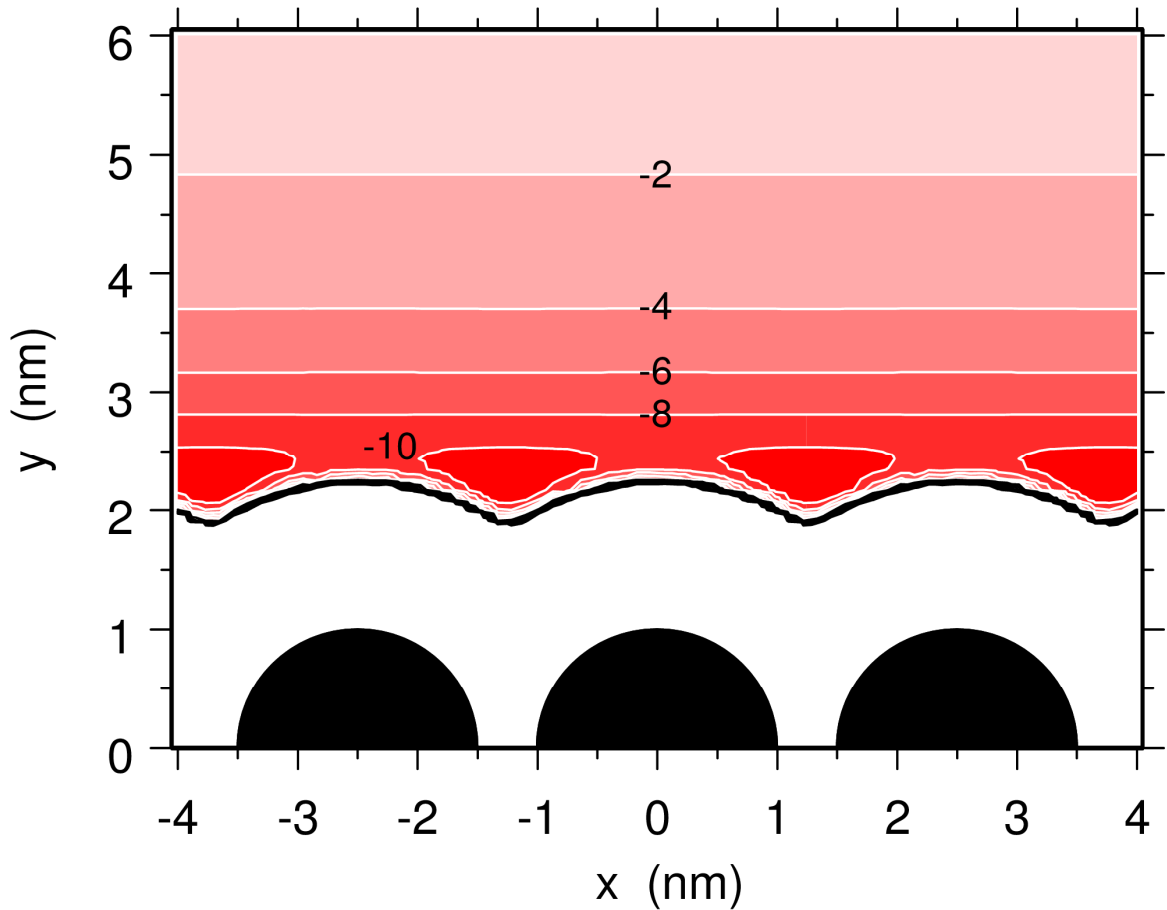

**Figure S2** : Potential energy felt by a protein chain aligned along the *y* axis when approaching an infinite DNA chain elongated along the *x* axis, for an effective valency Z=1.37. Both chains are at equilibrium with respect to their stretching and bending degrees of freedom. The black disks represent DNA beads. In this geometry the potential has rotational symmetry along the *x* axis. (*x,y*) denote the coordinates of the center of the terminal bead of the vertical protein chain that lies closest to the horizontal DNA chain. The minima of the potential energy surface ($-12.0\,k_\text{B}T$) are located at equal distances from two successive DNA beads. Contour lines are separated by $2\,k_\text{B}T$.

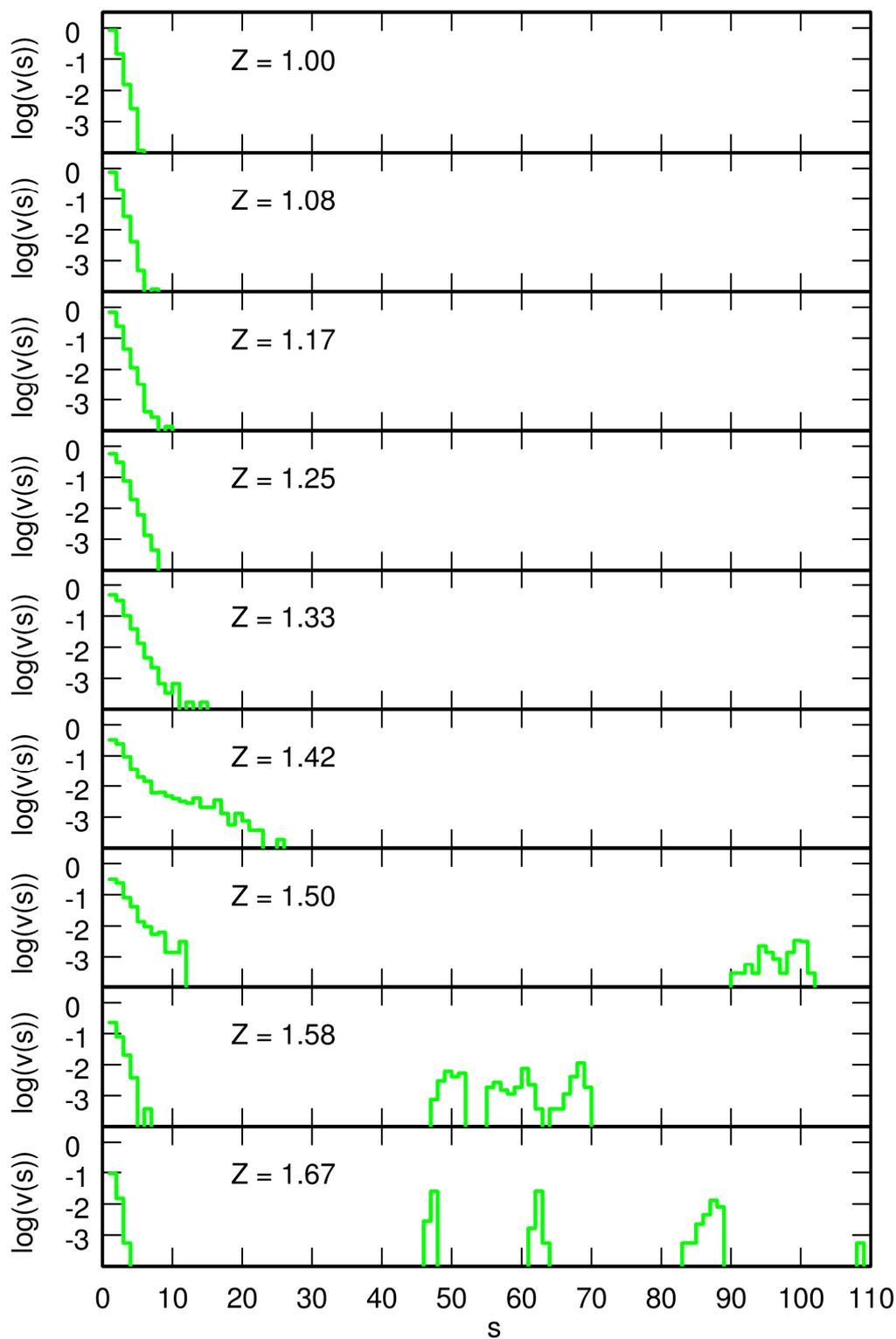

**Figure S3**: Decimal logarithm of $v(s)$, the probability distribution for a protein cluster to contain $s$ protein chains, for values of $Z$ increasing from 1.00 to 1.67. Each plot was obtained from a single equilibrated simulation with 200 protein chains by averaging over a time interval of 0.1 ms.

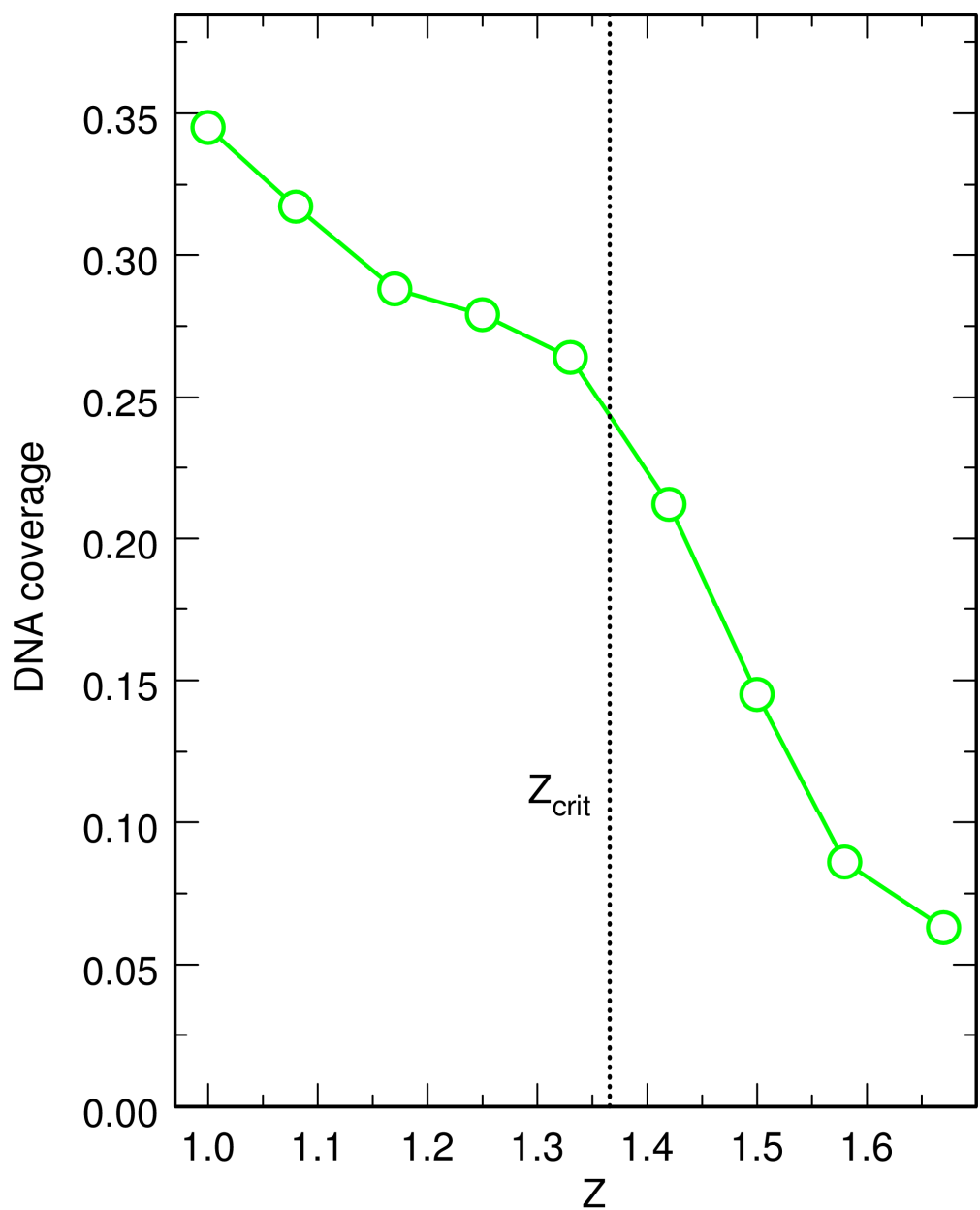

**Figure S4**: Plot, as a function of Z, of the fraction of the DNA chain covered by protein chains, for 200 protein chains. A bead $k$ of the DNA chain is considered to be "covered" by a protein chain if it is bound to a protein chain or surrounded, in the range $[k-5, k+5]$, by two beads that are bound to protein chains belonging to the same cluster.